\documentclass[11pt,a4paper]{article}

\usepackage{amsmath,amsfonts,amsbsy,amssymb,array,accents,dsfont}
\usepackage{enumerate,array,latexsym,graphicx,mathrsfs,verbatim,psfrag}
\usepackage{bm} 
\usepackage{bbm,braket}
\usepackage[normalem]{ulem}
\usepackage{booktabs} 
\usepackage[usenames]{color}

\usepackage[UKenglish]{babel}
\usepackage{fancybox}

\usepackage[nosort]{cite}
\usepackage{chngpage} 

\usepackage{enumitem}
\setlist{itemsep=0pt}

\usepackage[colorlinks=true,      linkcolor=darkblue,      urlcolor=darkblue,      
            filecolor=darkblue,      citecolor=darkblue,       pdfstartview=FitH,     
						pdfpagemode=UseNone,      bookmarksopen=true]{hyperref}  
\usepackage[all]{hypcap}     

\def\eq#1{(\ref{#1})}

\newcommand{\captionfonts}{\small}
\makeatletter  
\long\def\@makecaption#1#2{%
  \vskip\abovecaptionskip
  \sbox\@tempboxa{{\captionfonts #1: #2}}%
 \ifdim \wd\@tempboxa >\hsize
    {\captionfonts #1: #2\par}
  \else
    \hbox to\hsize{\hfil\box\@tempboxa\hfil}%
  \fi
  \vskip\belowcaptionskip}
\makeatother   

\DeclareMathSymbol{\medhatsym}{\mathord}{largesymbols}{"62} 

\DeclareMathSymbol{\medtildesym}{\mathord}{largesymbols}{"65}
\newcommand\lowermedtildesym{
  \text{\smash{\raisebox{-1.35ex}{%
    $\medtildesym$}}}}
\newcommand\medtilde[1]{
  \mathchoice
    {\accentset{\displaystyle\lowermedtildesym}{#1}}
    {\accentset{\textstyle\lowermedtildesym}{#1}}
    {\accentset{\scriptstyle\lowermedtildesym}{#1}}
    {\accentset{\scriptscriptstyle\lowermedtildesym}{#1}}
}

\def\({\left(}
\def\){\right)}
\def\[{\left[}
\def\]{\right]}

\def\barray{\begin{array}}
\def\earray{\end{array}}
\def\be{\begin{equation}}
\def\ee{\end{equation}}
\def\bea{\begin{eqnarray}}
\def\eea{\end{eqnarray}}
\def\bal{\begin{align}}
\def\eal{\end{align}}

\mathchardef\mhyphen="2D

\newcommand{\tr}{\mathrm{Tr}\;\!}

\numberwithin{equation}{section} %

\makeatletter
\g@addto@macro\bfseries{\boldmath}
\makeatother

\definecolor{cardinal}{rgb}{0.6,0,0}
\definecolor{darkgreen}{rgb}{0,0.4,0}
\definecolor{purple}{rgb}{0.5, 0, 0.5}
\definecolor{golden}{rgb}{0.92, 0.7, 0}
\definecolor{midnight}{rgb}{0, 0, 0.5}
\definecolor{darkblue}{rgb}{0, 0, 0.8}

\definecolor{emeraude}{RGB}{34,120,15}
\definecolor{turquoise}{RGB}{49, 140, 231}
\definecolor{framboise}{RGB}{199, 44, 72}

\def\cG{{\cal G}}

\def\cJ{{\cal J}}

\def\cM{{\cal M}}
\def\cN{{\cal N}}
\def\cO{{\cal O}}

\def\sst#1{\scriptscriptstyle{#1}}

\begin{document}

\begin{flushright}
%
%
\end{flushright}

\vspace{18mm}

\begin{center}

{\huge \bf{Four-point correlators in $\mathcal{N}=4$ SYM \vspace{5mm}\\ from AdS$_5$ bubbling geometries}}

\vspace{22mm}

{\large
\textsc{David Turton,\;\,Alexander Tyukov}}

\vspace{15mm}

\baselineskip=16pt
\parskip=3pt

Mathematical Sciences and STAG Research Centre,\\ University of Southampton,\\ Highfield,
Southampton SO17 1BJ, UK \\

\vspace{12mm}

{\upshape\ttfamily d.j.turton @ soton.ac.uk, a.tyukov @ soton.ac.uk} \\

\vspace{23mm}

\textsc{Abstract}

\end{center}

\begin{adjustwidth}{10mm}{10mm} 
%
\vspace{3mm}
\baselineskip=14.5pt
\noindent
Four-point correlation functions are observables of significant interest in holographic field theories.
We compute an infinite family of four-point correlation functions of operators in short multiplets of 4D $\mathcal{N}=4$ super Yang-Mills theory in the supergravity regime, by studying the quadratic fluctuations around non-trivial supergravity backgrounds.
The supergravity backgrounds are supersymmetric smooth geometries in the family derived by Lin, Lunin and Maldacena. 
The light probes comprise an infinite sequence of Kaluza-Klein harmonics of the dilaton/axion.
For generic parameter values, the supergravity backgrounds are dual to heavy CFT states.
However we focus on the limit in which the dual CFT states become light single-particle states.
The resulting all-light four-point correlators are related by superconformal Ward identities to previously known four-point correlators of half-BPS chiral primary operators.
By verifying that the Ward identities are satisfied, we confirm the validity of the supergravity method.

\end{adjustwidth}

\thispagestyle{empty}

\newpage


%
%


\baselineskip=15pt
\parskip=3pt


\section{Introduction}

Holographic duality has been a major theme in fundamental theoretical physics over more than 25 years. The most studied duality is the one proposed between type IIB string theory on AdS$_5\times $S$^5$ and 4D $\cN=4$ super Yang-Mills (SYM) theory with gauge group SU($N$)~\cite{Maldacena:1997re}. In the limit of large $N$ and large 't Hooft coupling, the string theory reduces to (type IIB) supergravity. Holographic observables computed in supergravity yield highly non-trivial predictions about the gauge theory in a regime of strong coupling.

Substantial supporting evidence for the duality has been obtained by studying protected quantities. Such quantities include two- and three-point correlators of half-BPS local operators in short multiplets containing chiral primary operators (CPOs)~\cite{Gubser:1998bc,Witten:1998qj,Freedman:1998tz,Lee:1998bxa,Petkou:1999fv,Baggio:2012rr}, as well as extremal and next-to-extremal four-point correlators of such operators, see e.g.~the review~\cite{DHoker:2002nbb}.
By contrast, generic four-point correlation functions are not protected, and contain dynamical information.

In the supergravity regime, four-point correlators of single-trace CPOs were first computed in progressively larger sub-families using tree-level Witten diagrams~\cite{Arutyunov:2000py,Arutyunov:2002fh,Berdichevsky:2007xd,Uruchurtu:2008kp,Uruchurtu:2011wh}. Later, a conjecture for the general such correlator was put forward~\cite{Rastelli:2016nze,Rastelli:2017udc} using Mellin space techniques~\cite{Mack:2009mi,Penedones:2010ue,Fitzpatrick:2011ia}. Soon thereafter, a much larger family of correlators was computed via Witten diagrams~\cite{Arutyunov:2018tvn}, confirming the conjecture in all examples considered. 
Using an analytic bootstrap approach, a generating function for general supergravity four-point amplitudes of operators in short multiplets was conjectured and verified in many 
examples~\cite{Caron-Huot:2018kta}; 
it was furthermore shown that this generating function implies the proposed Mellin space formula of~\cite{Rastelli:2016nze,Rastelli:2017udc}.

More recently, there has been much interest in computing integrated four-point correlators, see e.g.~\cite{Binder:2019jwn,Dorigoni:2021bvj,
Brown:2023zbr,Alday:2023pet}, 
as well as correlators in a certain large-charge 't Hooft limit, both for integrated and unintegrated correlators; see e.g.~\cite{Paul:2023rka,Caetano:2023zwe,Brown:2024yvt} and refs within. 
For a recent review on four-point correlators of half-BPS local operators,
see~\cite{Heslop:2022xgp}.

Separately, in AdS$_3$/CFT$_2$ holography, over the past several years there has been significant progress in computing four-point correlators, via the method of computing two-point functions of massless supergravity fields probing families of smooth supergravity solutions that generically correspond to heavy multi-trace CFT states. Several families of heavy-heavy-light-light (HHLL) correlators have been computed in this way~\cite{Galliani:2017jlg,Bombini:2017sge,Bombini:2019vnc,Giusto:2023awo}, 
where the two heavy states correspond to the supergravity solution and the two light operators correspond to the probe fluctuations. 
In the present work, `heavy' shall denote CFT operators that have conformal dimensions that scale linearly with the central charge $c$ in the large $c$ limit, while `light' shall denote CFT operators whose scaling dimensions are independent of $c$ in the large $c$ limit.

Moreover, by taking the limit in which the heavy background becomes a light single-particle perturbation of the vacuum, all-light (LLLL) four-point correlators of single-trace operators were also computed by this method~\cite{Giusto:2018ovt,Giusto:2019pxc,Giusto:2020neo}; see also the related works~\cite{Bena:2019azk,Rastelli:2019gtj,Giusto:2020mup,Ceplak:2021wzz}.
In many (though not all) cases, the all-light limit of the AdS$_3$ supergravity HHLL correlators did not yield the full LLLL correlators, and  
some terms of those all-light correlators were fixed by consistency considerations.
The advantage of this method is that it bypasses the difficulties of Witten diagram computations. This is particularly important in AdS$_3$/CFT$_2$ holography; indeed, the correlators computed via this method were the first LLLL four-point correlators to be computed in AdS$_3$/CFT$_2$. 
In related work, using a family of worldsheet models describing strings probing NS5-F1-P backgrounds developed in~\cite{Martinec:2017ztd,Martinec:2018nco,Martinec:2019wzw,Martinec:2020gkv,Bufalini:2021ndn,Martinec:2022okx}, a family of HHLL correlators was shown to agree between worldsheet CFT, supergravity, and holographic CFT~\cite{Bufalini:2022wyp,Bufalini:2022wzu}.

These developments give strong motivation to investigate the computation of both HHLL and LLLL correlators in AdS$_5$/CFT$_4$ holography using light probes of smooth supergravity solutions. The natural arena to start such investigations is the half-BPS sector of the theory.
A well-known family of half-BPS asymptotically AdS$_5 \times$S$^5$ smooth horizonless supergravity solutions was derived by Lin, Lunin and Maldacena (LLM)~\cite{Lin:2004nb}. This family includes solutions describing various configurations, including supergravity waves, D-branes, and fully backreacted ``bubbling'' solutions with non-trivial topology. These bubbling solutions have been interpreted as microstates of singular half-BPS `superstar' solutions~\cite{Myers:2001aq} that can be thought of as incipient black holes, in the sense that any energy added above extremality produces a horizon~\cite{Balasubramanian:2005mg}.

In the holographically dual 4D $\cN=4$ SYM theory on $\mathbb{R}\times$S$^3$, the dual half-BPS states correspond to generically multi-trace operators built from a holomorphic combination of two of the six hermitian scalars, $Z = \phi_1 + i \phi_2$. These operators have conformal weight equal to a particular U(1) R-charge in the SO(6) R-symmetry group, $\Delta=J$~\cite{Lin:2004nb}.

In the present work, we focus on the class of LLM solutions that have the same topology as global AdS$_5\times$S$^5$. These describe backreacted supergraviton waves, and in this sense are analogous to 
AdS$_3\times$S$^3$ Lunin-Mathur solutions~\cite{Lunin:2001jy} with connected, non-intersecting profiles, and related (albeit less supersymmetric) solutions known as superstrata~\cite{Bena:2015bea,Bena:2016agb,Bena:2016ypk,Bena:2017upb,Tyukov:2017uig,Bena:2017xbt,Ceplak:2018pws,Heidmann:2019xrd,Ceplak:2022pep,Ceplak:2024dbj}.
Generically, these LLM solutions are dual to heavy states of the dual CFT, however in a specific limit they reduce to the linearised fields of a supergravity fluctuation.

In this paper we compute an infinite family of holographic four-point correlators by computing supergravity two-point correlators on the background of a set of LLM solutions. LLM solutions are characterised by colouring a two-plane into black and white regions, denoting which of two three-spheres shrinks smoothly at a given point in the two-plane. The total area of the black regions is fixed. The solutions we study have a single compact black region. In plane polar coordinates $(r,\tilde{\phi})$, the boundary of the black region of the LLM solutions we study is specified by the following set of profiles:
\begin{align}
\label{eq:llm-ripple-1}
r(\tilde\phi) \,=\;\! \sqrt{1+\alpha \cos(n \tilde\phi)} \;,
\end{align}
where $\alpha < 1$. For small $\alpha$, this profile describes a ripple on a circle; this configuration was studied in~\cite{Skenderis:2007yb}.  In the present work, we moreover focus on the profile with $n=2$, and we shall work in perturbation theory in small $\alpha$.

The light supergravity probes we consider are the IIB dilaton and/or axion, whose linearized fluctuations satisfy the 10D minimally coupled massless scalar wave equation on the curved background of interest. We expand all quantities in harmonics on S$^5$. We take the source to be a lowest-weight 
scalar spherical harmonic on S$^5$, which we denote by $Y^{(k,-k)}$. We work with general non-negative integer $k$. 

In the CFT, the holographic duals of these light probes are descendants of the single-particle CPOs $\cO_{k+2}\sim \tr Z^{k+2}+\cdots$ and their conjugates $\bar\cO_{k+2}$. 
Here the ellipses denote multi-trace admixtures 
\cite{Aprile:2020uxk}, however these 
will not play any role in the present work.
The particular descendants we consider are of the form $\mathcal{D}_k  \sim Q^4 \cO_{k+2}$ and their conjugates $\bar{\mathcal{D}}_k$, which have dimension $\Delta = k+4$ and $R$-charge $J=\pm k$ respectively.
The precise map between these CFT operators and the dilaton/axion modes in supergravity is determined by the `bonus' U(1) hypercharge symmetry \cite{Intriligator:1998ig}, see e.g.~\cite{Liu:1999kg,DHoker:2002nbb}.
For $k=0$ we have the zero-mode of the dilaton/axion, a particular combination of which corresponds to the descendant $\mathcal{D}_0$ of the lightest CPO, $\cO_{2} = \tr Z^2$, which lies in the stress-tensor multiplet of $\cN=4$ SYM.

From the supergravity solution specified by the LLM profile~\eqref{eq:llm-ripple-1}, one can read off the energy above the global AdS$_5 \times $S$^5$ vacuum and R-charge~\cite{Skenderis:2007yb},
\be
\label{eq:e-j-a}
E \;=\; J \;=\; \frac14 \alpha^2 N^2 \;.
\ee
The precise form of the heavy CFT state dual to the LLM geometry with profile \eqref{eq:llm-ripple-1} is not known for general $\alpha$. It has been proposed that the CFT state involves a coherent superposition of many powers of the operator $\cO_2 = \tr Z^2$~\cite{Skenderis:2007yb}, however it was recently argued that this proposal must be modified to include one or more other single-trace operators~\cite{Giusto:2024trt}.

By contrast, at linear order in small $\alpha$, the holographically dual CFT state is known precisely, and this will suffice for the present work. The linearised supergravity fields are among the fluctuations in global AdS$_5$ classified in~\cite{Kim:1985ez}, as shown in~\cite{Grant:2005qc}.
These fields are dual to 
the state $\ket{\cO_2}$ corresponding to 
the chiral primary operator $\cO_2=\tr Z^2$. That is, the expansion of the dual heavy state $\ket{H_\alpha}$ to order $\alpha$ is
\be
\label{eq:H-to-order-alpha}
\ket{H_\alpha}~=~ \ket{0} + \alpha\ket{\cO_2} + \ldots \,,
\ee
where $\ket{0}$ is the conformal vacuum and where the ellipses denote terms at order $\alpha^2$ and higher.

In the perturbative approach we pursue, the first non-trivial  holographic four-point correlator arises at order $\alpha^2$. To derive this correlator, we compute the background at order $\alpha^2$ in closed form, and construct a diffeomorphism that puts it into Kaluza-Klein form and into de Donder-Lorentz gauge.

When $\alpha$ is small compared to 1 but independent of $N$, from Eq.\;\eqref{eq:e-j-a} the state can be described as ``perturbatively heavy'', see e.g.~\cite{Balasubramanian:2017fan}. By contrast, when $\alpha$ is of order $1/N$, the state becomes light. The supergravity calculation we perform
is insensitive to this distinction; for each value of $k$, there is a unique dynamical four-point supergravity correlator at order $\alpha^2$, which is both the leading dynamical term in the perturbative expansion of a HHLL correlator, and an LLLL correlator (c.f.~the discussion in~\cite{Ceplak:2021wzz}).

Using this fact, we focus our analysis of the supergravity correlator on the light limit. In the CFT, from Eq.~\eqref{eq:H-to-order-alpha} the order $\alpha^2$ dynamical part of the HHLL correlators $\langle H_\alpha(0) \bar{H}_\alpha(\infty) \bar{\mathcal{D}}_{k}(\vec{n}) \mathcal{D}_{k} (\vec{x})\rangle$ is the dynamical part of the following family of LLLL correlators,
\be
\label{eq:sugra-llll-desc-1}
\langle \cO_2(0) \bar{\cO}_2(\infty)  \bar{\mathcal{D}}_{k} (\vec{n}) \mathcal{D}_{k} (\vec{x}) \rangle \,,
\ee
where $\vec{n}$ denotes the position of the source on the boundary.
%
The family of all-light correlators \eqref{eq:sugra-llll-desc-1} is related by a superconformal Ward identity to the family of correlators of single-trace chiral primaries 
$\langle \cO_2(0) \bar{\cO}_2(\infty) 
\bar{\cO}_{k+2}(\vec{n})  \cO_{k+2} (\vec{x}) \rangle$, 
which were computed in tree-level supergravity for $k=0$ in~\cite{Arutyunov:2000py}, and 
for general positive integer $k$ 
in~\cite{Uruchurtu:2008kp}. The 
Ward identities were analyzed in~\cite{Drummond:2006by,Goncalves:2014ffa}. 
We write the supergravity expression for the all-light correlators in Mellin space and  
verify that the Ward identities are satisfied for all $k$. This establishes the validity of the supergravity method, and that the light limit gives the full LLLL correlators.

The structure of this paper is as follows. In Section \ref{sec:backgd-fields} we review the family of LLM solutions, and present the background metric for the profile~\eqref{eq:llm-ripple-1} in closed form to order $\alpha^2$. In Section \ref{sec:probe-calc} we describe the supergravity calculation of the holographic correlators. In Section~\ref{sec:mellin-ward} we discuss the holographically dual CFT correlators in Mellin space, and verify the superconformal Ward identity. In Section~\ref{sec:disc} we discuss our results. Some technical details are given in two appendices.

{\bf Note:} We are aware of independent related work by F.~Aprile, S.~Giusto and R.~Russo
which studies a scalar perturbation of a different geometry (the one constructed in~\cite{Giusto:2024trt}), obtaining the $k=0$ correlator in the family \eqref{eq:sugra-llll-desc-1} and a generalisation with double-particle
states~\cite{Aprile:2024}.


\section{Supergravity background fields}
\label{sec:backgd-fields}

\subsection{LLM solutions in AdS$_5\times$S$^5$}

We begin by briefly recalling the general form of the asymptotically AdS$_5\times$S$^5$ LLM solutions~\cite{Lin:2004nb}. 
Only the metric and five-form field strength are excited, as follows:
\begin{align}
     ds^2 &\,=\, -h^2(dt+V_i dx^i)^2 + h^2(dy^2+dx^i dx^i) + y e^G d\Omega_3^2 +y e^{-G} d\medtilde{\Omega}_3^2 \,, \nonumber \\
    h^{-2} &\,=\, 2y \cosh{G}\,, \qquad z=\frac{1}{2} \tanh{G}\,,\nonumber \\ 
    y\partial_y V_i &\,=\, \epsilon_{ij} \partial_j z \,, \qquad y\left(\partial_i V_j - \partial_j V_i\right) = \epsilon_{ij} \partial_y z\,, \nonumber \\
   F_5 &\,=\, F_{\mu\nu} dx^{\mu} \wedge dx^{\nu} \wedge d\Omega_3 + \tilde{F}_{\mu\nu} dx^{\mu} \wedge dx^{\nu} \wedge d\medtilde{\Omega}_3\,, \nonumber \\
   F &\,=\, d B_t \wedge (dt+V) + B_t dV + d\hat{B}\,, \\
   \tilde{F} &\,=\, d \tilde{B}_t \wedge (dt+V) + \tilde{B}_t dV + d\tilde{B}\,, \nonumber \\
   B_t &\,=\, -\frac{1}{4} y^2 e^{2G}\,, \qquad  \tilde{B}_t = -\frac{1}{4} y^2 e^{-2G}\,,\nonumber  \\
   d\hat{B} &\,=\,  -\frac{1}{4} y^3 *_3 d\left(\frac{z+\frac{1}{2}}{y^2}\right)\,, \qquad  d\tilde{B} =  -\frac{1}{4} y^3 *_3 d\left(\frac{z-\frac{1}{2}}{y^2}\right)\,,
   \nonumber 
\end{align}
where $i=1,2$ and $*_3$ is the Hodge dual on $\mathbb{R}^3$ parameterized by $(y,x_1,x_2)$, and where
\begin{align}
   z(x_1,x_2,y) &\;=\; \frac{y^2}{\pi} \int_{R^2} \frac{z(x'_1,x'_2,0)dx'_1 dx'_2}{((x-x')^2+y^2)^2}\;,\\
   V_i(x_1,x_2,y) &\;=\; \frac{\epsilon_{ij}}{\pi} \int_{R^2} \frac{z(x'_1,x'_2,0) (x_j-x'_j) dx'_1 dx'_2}{((x-x')^2+y^2)^2}\;.
\end{align}
In the coloring of the $\mathbb{R}^2$ at $y=0$ mentioned in the Introduction, the black regions denote the boundary condition $z=+1/2$ (in our conventions), and are interpreted as `droplets' in a free fermion phase space, which is related holographically to the free fermion description of the corresponding operators in the CFT~\cite{Corley:2001zk,Berenstein:2004kk}. In the black regions, the second sphere $\medtilde{\mathrm{S}}^3$ shrinks smoothly. By contrast, in the white regions, the boundary condition is $z=-1/2$, and instead the first sphere ${\mathrm{S}}^3$ shrinks smoothly.

As mentioned in the Introduction, 
we take the profile function to be given by
\begin{align}
\label{eq:ripplon-prof}
r(\tilde\phi) \,=\;\! \sqrt{1+\alpha \cos(n \tilde\phi)} \;,
\end{align}
where $\alpha<1$, and where $(r,\tilde\phi)$ are plane polar coordinates on $\mathbb{R}^2$ such that $x_1=r\cos\tilde{\phi}$ and $x_2=r\sin\tilde{\phi}$. When $\alpha$ is small, this profile describes a ripple deformation of a unit circle. 
While we focus on $n=2$ in this paper, we shall keep $n$ general in some of the following equations.

When $\alpha=0$, the profile is a circle, and the background is empty global AdS. The quantities $z$ and $V$ are then given by~\cite{Lin:2004nb} 
\begin{align}
    z &\,=\, -\frac{r^2+y^2-1}{2 \sqrt{(r^2+1+y^2)^2-4r^2}}\,,\nonumber\\
    V_{\tilde\phi} &\,=\, \frac{1}{2} \left(1-\frac{r^2+y^2+1}{\sqrt{(r^2+1+y^2)^2-4r^2}}\right), \qquad V_r \,=\, 0\,.
\end{align}
Upon making the following change of coordinates,
\be
y\,=\,R\cos\theta\,,\qquad r\,=\,\sqrt{R^2+1}\sin\theta\,,\qquad \tilde\phi \,=\, {\phi}-t\,,
\ee
the line element and five-form field strength take the form of empty global AdS$_5\times$S$^5$ (we use units in which the radii of AdS$_5$ and S$^5$ are set to 1), 
\be
\label{eq:metric0}
ds^2 \,=\,
-(R^2+1)dt^2 + \frac{dR^2}{R^2+1}+R^2 d\medtilde{\Omega}_3^2
+ d\theta^2 + \sin^2 \theta d\phi^2 + \cos^2\theta d\Omega_3^2 
\,,
\ee
\be
    F_5 \,=\, R^3\, dt\wedge dR \wedge d\medtilde{\Omega}_3 + \cos^3\theta\sin\theta\, d\theta \wedge d\phi \wedge d{\Omega}_3\,.
\ee
%

\subsection{First-order background}

We shall work perturbatively in small $\alpha$, to order $\alpha^2$. 
We thus expand the metric as
\be
     g =g^{(0)} + \alpha \:\! g^{(1)} +\alpha^2 \:\! g^{(2)} \,,
\ee
and similarly for the five-form field strength.

At order $\alpha$, the ripple of the LLM droplet corresponds to a linearised supergraviton in the order $\alpha^0$ background~\cite{Lin:2004nb}. The background fields that arise from the profile \eqref{eq:ripplon-prof} contain (negative) powers of the quantity
\be
\Sigma \,=\, R^2 + \cos^2\theta \;,
\ee
and are given in~\cite{Grant:2005qc}. It is convenient to make a linearised diffeomorphism to put the order $\alpha$ fields into the form of the fluctuation analysis of~\cite{Kim:1985ez}. 
We denote the vector field that generates this linearised diffeomorphism by $\xi^{(1)}$.
For $n=2$, the non-zero components of $\xi^{(1)}$ are~\cite{Grant:2005qc}
\begin{align}
    \xi^{(1)}_t &\,=\, -\frac{1}{2}  \frac{\sin ^2\theta}{R^2+1} \sin \!\!\:\big(2 (t-\phi)\big)\,,\nonumber\\
    \xi^{(1)}_R &\,=\, \frac{1}{2} \frac{\sin^2\theta}{\Sigma}  \frac{R}{R^2+1} \frac{\sin ^2\theta}{R^2+1} \cos \!\!\:\big(2 (t-\phi)\big)\,,\label{eq:xi-1}\\
    \xi^{(1)}_{\theta} &\,=\, \frac{1}{4} \frac{\sin 2\theta}{\Sigma} \frac{\sin ^2\theta}{R^2+1}  \cos \!\!\:\big(2 (t-\phi)\big)\,.
    \nonumber
\end{align}
%
Denoting AdS$_5$ indices by $\mu,\nu,\ldots$ and S$^5$ indices by $a,b,\ldots$, the resulting metric is in de Donder-Lorentz gauge, defined by
\be
    D^a h_{(ab)} \,=\, D^a h_{a\mu} \,=\, 0 \,,
\ee
where round brackets on indices denote symmetric traceless part.

The order $\alpha$ metric $g^{(1)}$ and four-form potential $A_4^{(1)}$ then take the form
\be
    g^{(1)}_{\mu\nu} = \sum_{n=\pm 2} \left(-\frac{6}{5} |n| s_n Y_n \, g^{(0)}_{\mu\nu} + \frac{4}{|n|+1} Y_n \nabla_{(\mu} \nabla_{\nu)} s_n \right), \qquad g^{(1)}_{\alpha\beta} = \sum_{n=\pm 2} 2|n|s_n Y_n \, g^{(0)}_{\alpha\beta}\,,
    \label{eq:g-1}
\ee
\be
    A_4^{(1)} = \sum_{n=\pm 2} \left( Y_n \star_{AdS_5} ds_n - s_n \star_{S^5} dY_n\right)\,,     
\ee
%
where the functions $s_n$ and $Y_n$ are given by 
\be
\label{eq:sn-Yn}
    s_n \,=\, \frac{|n|+1}{8|n|(R^2+1)^{|n|/2}} e^{i n t}\,, \qquad Y_n \,=\, e^{i n \phi} \sin^{|n|}\theta\,.
\ee
These are eigenfunctions of the Laplacians on AdS$_5$ and S$^5$ respectively, with eigenvalues given by
\be
    \square_{\mathrm{\sst A}} s_n = n(n-4)s_n\,, \qquad \square_{\mathrm{\sst S}} Y_n = -n(n+4)Y_n\,.
\ee

\subsection{Second-order background}

At order $\alpha^2$, we compute the closed-form background fields that follow from the profile~\eqref{eq:ripplon-prof}. The expressions for the fields that result directly from this computation are lengthy and not particularly illuminating, so we shall not write them explicitly. The important point is that the fields again contain (negative) powers of $\Sigma$. So it is again convenient to make a diffeomorphism to put the second-order metric $g^{(2)}$ into Kaluza-Klein form, and into de Donder-Lorentz gauge. To do so, on top of the second-order effects of the diffeomorphism generated by $\xi^{(1)}$, we make an additional order $\alpha^2$ diffeomorphism along the integral curves of another vector field $\xi^{(2)}$. We find that $\xi^{(2)}$ has the following non-zero components: 
\begin{align}
\label{eq:gauge-2}
    \xi^{(2)}_t &\,=\, -\frac{\left(1-4 R^2 +\cos 2 \theta \right) \sin^4\theta \sin \big(4 (t-\phi)\big)}{16 \Sigma}\,, \nonumber\\
    \xi^{(2)}_R &\,=\, \frac{1}{64} R \Bigg(-\frac{\left(3+R^2\right) (\cos 4 \theta -8 \cos 2 \theta)}{\left(R^2+1\right)^3}
    \cr
    &{}\qquad\qquad~~
    -\frac{12\left(1 + 2 R^2 +2 R^4+\left(2 R^2+1\right) \cos 2 \theta\right)}{\Sigma^2} 
    +\frac{14 \sin ^4\theta  \cos \big(4 (t-\phi)\big)}{\left(R^2+1\right)^2}
    \\
    &{}\qquad\qquad~~
    -\frac{  \left(25+24 R^2 +56 R^4 + 3 \cos 4 \theta +4 \left(7+22 R^2\right) \cos 2 \theta \right) \sin^4\theta \cos \big(4 (t-\phi)\big)}{4\left(R^2+1\right)^2 \Sigma^2}\Bigg)\,,\nonumber\\
    \xi^{(2)}_{\theta} &\,=\,\frac{1}{12}\sin 2\theta \left( \frac{\left(1+3 R^2\right)\cos ^2\theta}{\left(R^2+1\right)^3}-\frac{9\cos ^2\theta}{\Sigma^2} -\frac{3 \left(1 -4 R^2 +\cos 2 \theta \right) \sin ^4\theta  \cos \big(4 (t-\phi)\big)}{2\Sigma^2}\right) \,. 
    \nonumber
\end{align}
The resulting second-order metric $g^{(2)}$ is still somewhat lengthy; its form is recorded in Appendix \ref{app:metric2}. The five-form flux transforms analogously.

We then expand $g^{(2)}$ in S$^5$ spherical harmonics, as follows.
In the expressions below, $\bar{g}_{\mu\nu}$ and $\bar{g}_{ab}$ denote the relevant components of the $\alpha^0$ vacuum solution $g^{(0)}$, and $h$ denotes general metric fluctuations.
\begin{align}
    g_{\mu\nu} &\,=\, \bar{g}_{\mu\nu} + h_{\mu\nu}\,, \qquad g_{\mu a} = h_{\mu a}\,,\\
    g_{ab} &\,=\, \bar{g}_{ab} + h_{ab}\,,\\
    h_{\mu\nu}  &\,=\, h'_{\mu\nu} -\frac{1}{3} \bar{g}_{\mu\nu} h^a_a\,. 
\end{align}
We expand in scalar, vector and tensor S$^5$ harmonics $Y^{I_1}$, $Y^{I_5}_a$, $Y^{I_{14}}_{(ab)}$ as~\cite{Kim:1985ez,Skenderis:2007yb}
\begin{align}
    h'_{\mu\nu}(x,y) &\,=\, \sum \tilde{h}_{\mu\nu}^{I_1}(x) Y^{I_1}(y)\,,\\
    h_{\mu a}(x,y) &\,=\, \sum \tilde{B}_{(v)\mu}^{I_5}(x) Y_a^{I_5}(y)\,,\\
    h_{ab}(x,y) &\,=\, \sum \hat{\phi}_{(t)}^{I_{14}}(x) Y_{(ab)}^{I_{14}}(y)\,,\\
    h^a_a(x,y) &\,=\, \sum \tilde{\pi}^{I_1}(x) Y^{I_1}(y)\,.
\end{align}
The same can be done for the five-form flux. However, since we focus on dilaton/axion probes, and since these do not couple to the five-form, we will not describe the details here.


\section{Massless scalar wave equation and its solution}
\label{sec:probe-calc}

\subsection{Perturbative equations and boundary condition}

The equation of motion for the linearized fluctuation of the dilaton/axion is the 10D minimally coupled massless scalar wave equation on the curved background of interest, which we write as
\be
\label{eq:box-phi-full}
    \square \Phi = 0\,.
\ee
We expand the d'Alembertian and the scalar field $\Phi$ perturbatively, using the notation
\begin{align}
     \Phi &\;=\;\Phi^{(0)} + \alpha \;\! \Phi^{(1)} +\alpha^2 \:\!\Phi^{(2)} \,,\\
     \square &\;=\; \square^{(0)} + \alpha \;\!\square^{(1)} +\alpha^2 \:\! \square^{(2)} \,.
\end{align}
The equation of motion \eqref{eq:box-phi-full} expands in the standard way, as
\begin{align}
    \square^{(0)} \Phi^{(0)} &\,=\, 0\,,\label{eq:box-phi-0}\\
    \square^{(0)} \Phi^{(1)} &\,=\, -\square^{(1)}\Phi^{(0)}\,,\label{eq:box-phi-1}\\
    \square^{(0)} \Phi^{(2)} &\,=\, -\square^{(2)}\Phi^{(0)}-\square^{(1)}\Phi^{(1)}\,.
    \label{eq:box-phi-2}
\end{align}
We moreover expand the scalar field $\Phi$ in scalar spherical harmonics on S$^{(5)}$, 
\be
\label{eq:B-I}
    \Phi \,=\, \sum\limits_{I_1} B^{I_1} Y^{I_1} \,,  
\ee
and expand the coefficients perturbatively in $\alpha$ as
\be
\label{eq:B-I-pert}
   B^{I_1} \,=\, B^{(0)I_1} + \alpha \:\! B^{(1)I_1} +\alpha^2 \:\! B^{(2)I_1} \,.
\ee
The boundary condition appropriate for the correlator of interest is
\be
\label{eq:bc}
    \lim_{R\to \infty} \Phi  \,=\, \frac{\delta(\vec{x}-\vec{n}) Y^{I}(y)}{R^{d-\Delta}} + \frac{b^{I}(\vec{x}) Y^{I}(y)}{R^{\Delta}} +\cdots\,,
\ee
where $d=4$ for AdS$_5$, $\vec{n}$ denotes the position of the source on the boundary, $b^I(\vec{x})$ denotes the response in the harmonic $Y^I$, and where the dots stand for response terms proportional to harmonics other than $Y^{I}$.

The response $b^{I}(\vec{x})$ arises at order $\alpha^2$. 
As mentioned in the introduction, it encodes the dynamical order $\alpha^2$ term in a heavy-light (HHLL) correlator, as well as the dynamical part of the following LLLL four-point correlator in the supergravity regime,
\be
b^{I}(\vec{x}) \;=\; 
    \big\langle \mathcal{O}_2 (0) \bar{\mathcal{O}}_2 (\infty) \bar{\mathcal{D}}_{k} (\vec{n}) \mathcal{D}_{k} (\vec{x}) 
    \big\rangle 
    \;.
\ee
We note that at the level of the supergravity calculation, there is no distinction between dilaton/axion probes. When interpreting the supergravity correlator in the CFT, as mentioned in the Introduction, we shall focus on the dual CFT operators that are U(1) hypercharge eigenstates, namely $\mathcal{D}_k  \sim Q^4 \cO_{k+2}$ and $\bar{\mathcal{D}}_k  \sim \bar{Q}^4 \bar\cO_{k+2}$, with dimension $\Delta = k+4$ and $R$-charge $J=\pm k$ respectively.

\subsection{Solution method and second-order source terms}

We now describe the iterative solution in brief outline, before presenting the details.
At order $\alpha^0$, as input we take the solution for $\Phi^{(0)}$ that corresponds to the descendant $\bar{\mathcal{D}}_k  \sim \bar{Q}^4 \bar\cO_{k+2}$, i.e.
\be
  \Phi^{(0)} \;=\; B^{(0)}(x) Y_{-k}(y)\;,
\ee
where we suppress SO(6) labels on $B^{(0)}$, and where $Y_{-k}(y) = Y^{(k,-k)}(y)$ is the lowest-weight harmonic on S$^5$ introduced in~\eqref{eq:sn-Yn}. $B^{(0)}(x)$ will be given explicitly in Eqs.~\eqref{eq:phi-0}--\eqref{eq:bhat-1}.

We next obtain the general solution to the order $\alpha$ equation of motion, Eq.~\eqref{eq:box-phi-1}. 
The SO(4) invariance of the background implies that only the subset of $S^5$ harmonics that are SO(4)-invariant appear. 
Since the background at order $\alpha$ has charge $\pm 2$, (c.f.~Eq.~\eqref{eq:g-1}), we obtain the following harmonic expansion,
\be
  \Phi^{(1)} \;=\; \sum_l B_{l,+}^{(1)}(x) Y^{(l,-k+2)}(y)+B_{l,-}^{(1)}(x) Y^{(l,-k-2)}(y) \;,
\ee
where $Y^{(l,m)}$ is a generic SO(4)-singlet harmonic on $S^5$.    

We then evaluate the sources on the right-hand side of the order $\alpha^2$ equation of motion~\eqref{eq:box-phi-2}. Since we are interested in the response in the harmonic $Y_{-k}(y)$, we project back on the lowest-weight harmonic before solving this equation. The solution at order $\alpha^2$ then has the form
\be
 \Phi^{(2)} \;=\; B^{(2)}(x) Y_{-k}(y)\;,
\ee
where we again suppress SO(6) labels on $B^{(2)}$.
Taking the limit of $B^{(2)}$ to the boundary of AdS$_5$, one reads off $b^I(\vec{x})\equiv b_k(\vec{x})$ in \eqref{eq:bc}.

We now describe the solution in more detail.
Solving the order $\alpha^0$ equation \eqref{eq:box-phi-0} subject to the boundary condition \eqref{eq:bc}, we obtain the appropriate bulk-to-boundary propagator,
\be
\label{eq:phi-0}
\Phi^{(0)} \;=\;
    \hat{B}_{k+4}(x) Y_{-k}(y)\,,
\ee
where $\hat{B}_{\Delta}$ denotes  the bulk-to-boundary propagator of dimension $\Delta$ with boundary point $\vec{n}$.

Without loss of generality, we introduce coordinates on the S$^3$ inside AdS$_5$ that are adapted to the specified boundary point $\vec{n}\in \mathbb{R}\times$S$^3$. 
Concretely, we use hyperspherical coordinates $(\beta,\gamma,\delta)$ 
on the three-sphere in the AdS$_5$ part of the 
order $\alpha^0$ background metric 
\eqref{eq:metric0},
\begin{align}
d\medtilde\Omega_3^2 \,&=\, d\beta^2 +\sin\beta^2 ( d\gamma^2 + \sin\gamma^2 d\delta^2 ) \;,
\end{align}
such that $\vec{n}$ is at $t=0$ and at the North pole of the three-sphere, that is $t=\beta=0$.
Of course, the equations of the present subsection generalise straightforwardly to other coordinates.
Then the bulk-to-boundary propagator takes the form
\be
\label{eq:bhat-1}
\hat{B}_{\Delta}(\vec{R}) \,\equiv\, \left( \frac12
\frac{1}{\sqrt{R^2+1}\cos t - R \cos \beta} \right)^\Delta \,.
\ee
We next solve for $\Phi^{(1)}$ as follows. First, we employ the additional gauge transformation found in \cite{Giusto:2024trt} which sets to zero $g^{(1)}_{tR}$ at the cost of introducing a non-zero $g^{(1)}_{\theta\phi}$.
In those coordinates, the source on the RHS of the order $\alpha$ equation of motion \eqref{eq:box-phi-1} simplifies considerably. We find the following set of solutions. For $k=0$, the source is zero, so the solution is trivial:
\be
    \Phi^{(1)'}_{k=0} \,=\, 0\,.
\ee
For $k=1$, we obtain
\be
\label{eq:phi1pk1}
    \Phi^{(1)'}_{k=1} \,=\, \hat{B}_{4}(x) s_{1}(x) Y_{1}(y)\,,
\ee
where $Y_{1}(y) = Y^{(1,1)}(y) = e^{i\phi}\sin\theta$ and $s_{1}(x)$ is given in~\eqref{eq:sn-Yn}. 

For $k\ge 2$, we obtain
\be
\label{eq:phi1pkge2}
    \Phi^{(1)'}_{k\ge 2} \,=\, - \frac{2k(k-1)}{k+3} \hat{B}_{k+3}(x) s_{1}(x) Y^{(k,2-k)}(y)\,.
\ee
We note that for $k\ge 2$ we have 
\be
2(k-1) Y^{(k,2-k)}(y) \,=\, e^{-i (k-2) \phi } \sin^{k-2}\theta  \Big( 
(k+1) \cos (2 \theta )+k-3
\Big).
\ee
Upon substituting this expression into the previous equation~\eqref{eq:phi1pkge2}, the $k\to 1$ limit can be taken, and is in agreement with the $k=1$ expression in~\eqref{eq:phi1pk1}.

Transforming back to the de Donder-Lorentz gauge in which we work adds a contribution
\be
    \delta\Phi^{(1)} \,=\, \frac{2}{3}\Big( (Y_2 \nabla^{M}s_2-s_2 \nabla^{M}Y_2) + c.c. \Big)\partial_M \Phi^{(0)}  \,, 
\ee
so that the total solution at order $\alpha$ is given by 
\be
   \Phi^{(1)} \,=\, \Phi^{(1)'} +\delta\Phi^{(1)}\,.  
\ee
For convenience, to write the sources on the RHS of the order $\alpha^2$ equation \eqref{eq:box-phi-2}, we introduce the notation 
\be
\label{eq:sources-defn}
    \mathcal{J}_1 \,=\, \square^{(2)}\Phi^{(0)}\,, \qquad \mathcal{J}_2 \,=\, \square^{(1)}\Phi^{(1)}\,.  
\ee
Then the order $\alpha^2$ equation \eqref{eq:box-phi-2} becomes 
\be
\label{eq:box-phi-2-J}
    \square^{(0)}\Phi^{(2)} \,=\, - (\mathcal{J}_1+\mathcal{J}_2)\,.
\ee
To extract $B^{(2)}$, we project the sources $\mathcal{J}_1$, $\mathcal{J}_2$
on the highest-weight spherical harmonic $Y_k$,
\be
\big\langle\mathcal{J}_{i}\big\rangle
\;\equiv \;
 \frac{1}{||Y_{k}||^2}  \int d\Omega_5 \,
    \mathcal{J}_i
    \, Y_{-k}^{*} \,,
    \qquad\quad i=1,2 \,.
\ee
Projecting similarly Eq.~\eqref{eq:box-phi-2-J}, we obtain the following five-dimensional equation to be solved,
\begin{align}
    \square_5^{(0)} B^{(2)} -m^2 B^{(2)}
    &{}\;=\;-\big\langle\mathcal{J}_1\big\rangle -\big\langle\mathcal{J}_2\big\rangle \;,
    \label{eq:proj-sources}
\end{align}
where $\square_5^{(0)}$ denotes the scalar Laplacian on global AdS$_5$.

The projected sources $\big\langle\mathcal{J}_1\big\rangle$ and $\big\langle\mathcal{J}_2\big\rangle$ depend only on $(t,R,\beta)$.
They  
involve linear combinations of $\hat{B}_{k+4}$, $\hat{B}_{k+3}$, and their derivatives, with coefficients depending on $k$ and $R$. 
For later convenience we choose to eliminate second $\beta$ derivatives using $\square^{(0)}\Phi^{(0)}=0$. It happens that this eliminates also all first derivatives with respect to $\beta$, leaving only $t$ and $R$ derivatives. 
The expressions for the projected sources are somewhat lengthy, and are recorded in Appendix \ref{app:sources}.

\subsection{Rewriting the second-order source terms}

Having computed the source terms on the right-hand side of the order $\alpha^2$ equation of motion, Eq.~\eqref{eq:proj-sources}, the next step is to solve it using the bulk-to-bulk propagator $G^{\mathrm{Glob}}_{\Delta}(\vec{R}'|\vec{R})$ 
(see e.g.~\cite[Eq.\;(6.12)]{DHoker:2002nbb}),
\be
\label{eq:corr-sug-1}
    B^{(2)} (\vec{R}) \;=\; 
    i \! \int d^5 \!\!\; \vec{R}' \sqrt{-g_{AdS_5}} \, G^{\mathrm{Glob}}_{\Delta}(\vec{R}'|\vec{R})
    \left(
\big\langle\mathcal{J}_1(\vec{R}')\big\rangle +\big\langle\mathcal{J}_2(\vec{R}')\big\rangle
    \right) .
\ee

To extract the four-point correlator, we examine the solution near the asymptotic boundary of AdS$_5$, as described in Eq.~\eq{eq:bc}. Upon doing so, the bulk-to-bulk propagator becomes a bulk-to-boundary propagator.
Concretely, at leading order in large $R$, the bulk-to-bulk propagator 
$G^{\mathrm{Glob}}_{\Delta}$ is related to the bulk-to-boundary propagator $K^{\mathrm{Glob}}_{\Delta}$ as follows: 
\be
    G^{\mathrm{Glob}}_{\Delta}(\vec{R'}|R,t,\vec{\Omega}) ~\;\to\;~ \frac{\Gamma(\Delta)}{2 \pi^{d/2} \Gamma(\Delta-\frac{d}{2}+1)} \frac{1}{R^{\Delta}} K^{\mathrm{Glob}}_{\Delta}(\vec{R'}|t,\vec{\Omega})\,,
\ee
where the bulk-to boundary propagator is given by
\be
\label{eq:k-glob-main}
    K^{\mathrm{Glob}}_{\Delta}(\vec{R'}|t,\vec{\Omega}) \;\equiv\; \left(\frac{1}{2}\frac{1}{\sqrt{1+R'^2} \, \cos(t-t')-R'\, \vec{\Omega}\cdot \vec{\Omega}'  }\right)^{\Delta}\,,
\ee
and where $\vec{\Omega}$ denotes a point on the three-sphere in AdS$_5$, represented as a unit vector in $\mathbb{R}^4$, such that\footnote{We shall not need similar notation for the three-sphere in S$^5$, so we do not use a tilde on $\vec{\Omega}$.}
\be
    \vec{\Omega}\cdot \vec{\Omega}' \;=\; \sin\beta \sin\beta' \left(\sin\gamma\sin\gamma'\cos(\delta-\delta')+\cos\gamma\cos\gamma'\right) + \cos\beta \cos\beta'\,.    
\ee
We thus obtain the response
\be
\label{eq:resp-hyp}
    b^{(2)}(t,\vec{\Omega}) \;=\;  {}  \frac{\Gamma(\Delta)}{2 \pi^{d/2} \Gamma(\Delta-\frac{d}{2}+1)} \!\; i \!\!\!\; \int d^5 \vec{R}' \sqrt{-g_{AdS_5}} \, K^{\mathrm{Glob}}_{\Delta}(\vec{R'}|t,\vec{\Omega}) \left(
\big\langle\mathcal{J}_1(\vec{R}')\big\rangle +\big\langle\mathcal{J}_2(\vec{R}')\big\rangle
    \right)
     \,.
\ee

We now rewrite the sum of the sources $\big\langle\mathcal{J}_1\big\rangle+\big\langle\mathcal{J}_2\big\rangle$ in terms of bulk-to-boundary propagators, in order to express the four-point correlators in terms of the usual $D$-functions that arise in holographic four-point correlators~\cite{DHoker:1999kzh}.
To do so, we Wick rotate
to Euclidean time $t_e=i t$, writing Euclidean
hyperspherical coordinates collectively as $\vec{R}_e=(t_e,R,\vec{\Omega})$, and define
\begin{align}
\begin{aligned}
\label{eq:b-to-bdy-hy}
B^+_\Delta(\vec{R}_e) \;&\equiv\; 
\left(\frac{e^{t_e}}{\sqrt{1+R^2}}\right)^{\Delta} ,
\\
B^-_\Delta(\vec{R}_e) \;&\equiv\; 
\left(\frac{e^{-t_e}}{\sqrt{1+R^2}}\right)^{\Delta},
\\
\hat{B}_\Delta(\vec{R}_e) \;&\equiv\; 
\left( \frac12
\frac{1}{\sqrt{R^2+1}\cosh t_e - R \cos \beta} \right)^\Delta \,,
\end{aligned}
\end{align}
where the last line is the Wick rotation of Eq.~\eqref{eq:bhat-1}.

When the total source $\langle\cJ_1\rangle+\langle\cJ_2\rangle$ is written in terms of these functions, the terms that contribute to the dynamical part of the four-point correlator of interest are those that involve a product of all three of $B^+_\Delta$, $B^-_\Delta$ and $\hat{B}_\Delta$, as will become clear shortly.
After discarding terms that do not contribute to the dynamical part of the correlator of interest, the resulting dynamical part of the total source, 
\be
\label{eq:j-dyna}
\mathcal{J} \;\equiv\; \Big(
\langle\cJ_1\rangle+\langle\cJ_2\rangle
\Big)\Big|_{\mathrm{dyna}}
\ee
is by construction a sum of products of all three of $B^+_\Delta$, $B^-_\Delta$ and $\hat{B}_\Delta$, with coefficients that are functions of $k$ only. The explicit expression for $\cJ$ is quite lengthy, and is recorded in Eq.~\eqref{eq:source-prop}.


\subsection{Four-point correlators in terms of D-functions}


We now transform to Euclidean Poincar\'e coordinates $\boldsymbol{w} \equiv (w_0,\vec{w})$ via
\begin{align}
    w_0 \,=\, \frac{e^{t_e} }{\sqrt{1+R^2}}\,, \qquad
    w_1 &\,=\, w_0 R \sin \beta \sin \gamma \cos \delta \,, \qquad
    w_2 \,=\, w_0 R \sin \beta \sin \gamma \sin \delta \,,\nonumber \\
    w_3 &\,=\, w_0 R \sin \beta \cos \gamma  \,,\qquad
    w_4 \,=\, w_0 R \cos \beta \,,
\end{align}
with line element
\begin{align}
    ds_{\mathrm{\sst{EAdS_5}}}^2 &\;=\;  \frac{1}{w_0^2}\bigg( dw_0^2 + \sum\limits_{i=1}^{4} dw_i^2 \bigg)\;.
\end{align}

From now onwards, we use $\vec{x}=(x^1,x^2,x^3,x^4)$ to denote the coordinates on the boundary $\mathbb{R}^4$ at $w_0=0$. The relation between these coordinates and those on the Euclidean cylinder $\mathbb{R}\times $S$^3$ at $R = \infty$ is
\be 
    x^1 = e^{t_e} \sin \beta \sin \gamma \cos \delta \,, ~~
    x^2 = e^{t_e} \sin \beta \sin \gamma \sin \delta \,, ~~
    x^3 = e^{t_e} \sin \beta \cos \gamma \,, ~~
    x^4 = e^{t_e} \cos \beta \,.
\ee

In Euclidean Poincar\'e coordinates, the bulk-to-boundary propagators with boundary points at $(w_0=0,\vec{x})$ and at $w_0=\infty$ are respectively~\cite{Witten:1998qj}
\begin{align}
\label{eq:k-ads5-poin}
K_\Delta(\boldsymbol{w}\:\!|\:\!\vec{x}) \,&\equiv\, \left(\frac{w_0}{w_0^2+|\vec{w}-\vec{x}|^2}\right)^{\Delta}\,,
\qquad 
K_\Delta(w_0\:\!|\:\!\infty) \,\equiv\, w_0^\Delta \;.
\end{align}
The bulk-to-boundary propagators in the two coordinate systems are related by
\be
    \label{eq:k-glob-k-poin}
    K^{\mathrm{Glob}}_{\Delta}(\vec{R}'_e|t_e,\vec{\Omega}) \,=\, |\vec{x}|^{\Delta} K_{\Delta}(\boldsymbol{w}|\vec{x})\,. 
\ee

We shall write the sources in terms of three bulk-to-boundary propagators, with respective boundary points at $t_e=\pm \infty$ and  the point $\vec{n}\in\mathbb{R}\times$S$^3$ defined by $\,t_e=\beta=0$.
The point at $t_e = \infty$ corresponds to $w_0 =\infty$, while $t_e \to -\infty$ maps to the origin $\vec{x}=0$ at $w_0=0$. The point $\vec{n}\in\mathbb{R}\times$S$^3$ 
maps to the point $\vec{x}=\vec{n}_x\equiv (0,0,0,1)$ at $w_0=0$.

Comparing expressions in hyperspherical and Poincar\'e coordinates, we have
\begin{align}
B^+_\Delta(\vec{R}_e) \,&=\, K_{\Delta}(\boldsymbol{w}|\infty) 
\,=\, \lim_{|\vec{x}|\to \infty}|\vec{x}|^{2\Delta}
K_{\Delta}(\boldsymbol{w}|\vec{x})
\,,
\nonumber\\
B^-_\Delta(\vec{R}_e) \,&=\, K_{\Delta}(\boldsymbol{w}|0) \,,
\nonumber\\
  \hat{B}_\Delta(\vec{R}_e) \,&=\, 
    K^{\mathrm{Glob}}_{\Delta}(\vec{R}'_e|\vec{n})
    \,=\,
    K_{\Delta}(\boldsymbol{w}|\vec{n}_x)\,,
\end{align}
where comparing to \eqref{eq:k-glob-k-poin} we note that $|\vec{n}_x |=1$.
Thus, each of $B^+_\Delta$, $B^-_\Delta$, $\hat{B}_\Delta$ maps to a bulk-to-boundary propagator in Euclidean Poincar\'e coordinates. 

The response $b^{(2)}$ in Eq.\;\eqref{eq:resp-hyp} encodes a four-point correlator on $\mathbb{R}\times$S$^3$. When we change to Poincar\'e coordinates, the boundary maps to $\mathbb{R}^4$ (plus the point at infinity).
The conformal transformation of a CFT operator of scaling dimension $\Delta$ from $\mathbb{R}\times$S$^3$ to flat $\mathbb{R}^4$ is 
\be
    O_{\Delta}^{\mathrm{cyl}} (t_e,\vec{\Omega}) = |\vec{x}|^{\Delta} O_{\Delta}^{\mathrm{flat}} (\vec{x})\,, 
\ee
where $|\vec{x}|
=e^{t_e}$.
Upon changing coordinates, the fourth bulk-to-boundary propagator changes similarly, see Eq.~\eqref{eq:k-glob-k-poin}, and we thus obtain the correlator on $\mathbb{R}^4$,
\be
 \langle \mathcal{O}_2 (0) \bar{\mathcal{O}}_2 (\infty) \bar{\mathcal{D}}_{k} (\vec{n}) \mathcal{D}_{k} (\vec{x}) \rangle\Big|_{\mathrm{dyna}} \;=\;  {} \frac{\Gamma(\Delta)}{2 \pi^{d/2} \Gamma(\Delta-\frac{d}{2}+1)} \int d^5 \boldsymbol{w'} \sqrt{\bar{g}} \, K_{\Delta}(\boldsymbol{w'}\:\! |\:\! \vec{x} ) \mathcal{J}(\boldsymbol{w'}) \;.
\ee

The correlator is now expressed as a sum of integrals over four bulk-to-boundary propagators, so it is natural to write it in terms of $D$-functions.
We recall that in $d$ boundary dimensions the $D$-functions $D_{\Delta_1 \Delta_2 \Delta_3 \Delta_4}(x_1, x_2, x_3, x_4)$ are defined by~\cite{DHoker:1999kzh}
\begin{align}
    D_{\Delta_1 \Delta_2 \Delta_3 \Delta_4}(x_1, x_2, x_3, x_4) &\,=\, \int d^{d+1} \boldsymbol{w} \sqrt{\bar{g}}\, \prod_{i=1}^4 K_{\Delta_i}(\boldsymbol{w}| \vec{x}) \\
    &\,=\, \Gamma\left(\frac{\hat{\Delta}-d}{2}\right) \frac{\pi^{d/2}}{2} \int_0^{\infty} \prod_{i=1}^4 \left[dt_i\frac{t_i^{\Delta_i-1}}{\Gamma(\Delta_i)}\right]\frac{1}{T^{\hat{\Delta}/2}} e^{-\sum_{i,j=1}^4 x_{ij}^2 \frac{t_i t_j}{2 T} }\,,  
    \nonumber
\end{align}
where $T\equiv\sum_i t_i$, $\hat{\Delta}\equiv\sum_i \Delta_i$, $x_{ij}\equiv x_i-x_j$, and $x_{ij}^2\equiv |x_{ij}|^2$.

We define the conformal cross-ratios $U$ and $V$ via
\be
    U \equiv \frac{x_{12}^2 x_{34}^2}{x_{13}^2 x_{24}^2}\,,\qquad
    V \equiv \frac{x_{14}^2 x_{23}^2}{x_{13}^2 x_{24}^2}\,.
\ee 
We also introduce the 
$\bar{D}$-functions $\bar{D}_{\Delta_1 \Delta_2 \Delta_3 \Delta_4}(U,V)$ via 
\be
  D_{\Delta_1 \Delta_2 \Delta_3 \Delta_4}(x_1, x_2, x_3, x_4) \,=\, \frac{\pi^{d/2}\Gamma\left(\frac{\hat{\Delta}-d}{2}\right)}{2\prod_{i=1}^4\Gamma(\Delta_i)} \frac{x_{14}^{\hat{\Delta}-2\Delta_1-2\Delta_4} x_{34}^{\hat{\Delta}-2\Delta_3-2\Delta_4}}{x_{13}^{\hat{\Delta}-2\Delta_4} x_{24}^{2\Delta_2}} \bar{D}_{\Delta_1 \Delta_2 \Delta_3 \Delta_4}(U,V)\,.  
\ee
This definition agrees with (5.7) in \cite{Dolan:2000ut} (we absorbed the factor of $\pi^{d/2}$ into $D$) and (A.5) in \cite{Giusto:2018ovt}.

We note that we can express $(U,V)$ in terms of $\mathbb{R}\times $S$^3$ coordinates on the boundary $\mathbb{R}^4$ by setting 
$x_1= 0$ ($t_e\to -\infty$), $x_2\to\infty$ ($t_e\to +\infty$), $x_3= \vec{n} = (0,0,0,1)$ ($t_e = 0$ and $\beta=0$), and $x_4^i$ being generic.   
This results in 
\be 
    U = x_{34}^2 = 1 + e^{2t_e} - 2e^{t_e} \cos \beta \,,
    \qquad V = x_4^2 = e^{2t_e}\,. 
\ee
We again see that the correlator has no dependence on the $S^2$ directions of $S^3$.
One can also parametrize the cross-ratios with the complex variable $z$, as is often done (see e.g.~\cite{Dolan:2006ec})\footnote{Note that many papers use conventions in which $z$ and $1-z$ are interchanged compared to Eq.~\eqref{eq:U-V}.}, 
\be
\label{eq:U-V}
    U = (1-z)(1-\bar{z})\,,\qquad V = z\bar{z}\,,
\ee
where $z=e^{t_e+i\beta}$.

Finally, we define the 
$\hat{D}$-functions (for $d=4$) via
\be
\begin{split}
     \hat{D}_{\Delta_1 \Delta_2 \Delta_3 \Delta_4}(z,\bar{z}) &\,=\, \lim_{|\vec{x}_2|\to\infty} |\vec{x}_2|^{2\Delta_2} D_{\Delta_1 \Delta_2 \Delta_3 \Delta_4}(\vec{x_1}=0,\vec{x_2},\vec{x_3}=\vec{n},\vec{x}) \\
     &\,=\, \frac{\pi^2\, \Gamma\left(\frac{\hat{\Delta}}{2}-2\right)}{2\prod_{i=1}^4 \Gamma(\Delta_i)} |z|^{\hat{\Delta}-2\Delta_1-2\Delta_4} |1-z|^{\hat{\Delta}-2\Delta_3-2\Delta_4} \bar{D}_{\Delta_1 \Delta_2 \Delta_3 \Delta_4}(z,\bar{z})\,. 
\end{split}
\ee
The correlator will then be expressed in terms of $\hat{D}$ functions.

The resulting expression for the correlator in terms of $\hat{D}$ functions can be read off from~\eqref{eq:source-prop}. Due to multiple algebraic identities between different $D$-functions, it is not in simplest form. 
The more natural way to present and analyse the correlator is in Mellin space, which we do in the next section.  
%


\section{Correlators in Mellin space and Ward identity}
\label{sec:mellin-ward}

\subsection{Supergravity four-point correlators in Mellin space}

In order to write the supergravity correlators in Mellin space, we first introduce some notation. We consider CFT four-point functions of scalar operators $\cO_i$ of respective dimensions $\Delta_i$. We specialise throughout to correlators with $\Delta_1=\Delta_2$ and $\Delta_3=\Delta_4$. Conformal invariance implies that the correlator takes the form
\be
\langle
\cO_1(x_1)\cO_2(x_2)\cO_3(x_3)\cO_4(x_4)
\rangle 
\,=\,
\frac{1}{ (x_{12}^2)^{\Delta_1} (x_{34}^2)^{\Delta_3}}
\cG
(U,V) \,,
\ee
where the conformal cross-ratios $U,V$ were defined in Eq.~\eqref{eq:U-V}.

As usual, we separate the supergravity correlator  into free and dynamical parts (see e.g.~\cite{Aprile:2017xsp}),
\be
\cG^{(\mathrm{{sugra}})}(U,V) \,=\, 
\cG^{(\mathrm{{free}})}(U,V)
+\cG^{(\mathrm{{dyna}})}(U,V) \,,
\ee
where $\cG^{(\mathrm{{free}})}(U,V)$ denotes the value 
in free $\cN=4$ SYM.
We focus on the dynamical  part of the correlator, which is what we have derived in the previous section. 
The corresponding Mellin amplitude $\cM(s,t)$ is defined via the following integral transform. Introducing explicit subscripts $\cG^{(\mathrm{dyna})}(U,V)=\mathcal{G}^{(\mathrm{dyna})}_{\Delta_1,\:\!\Delta_3}(U,V)$, we have (we use the conventions of~\cite{Giusto:2019pxc})
\begin{align}
\mathcal{G}^{(\mathrm{dyna})}_{\Delta_1,\:\!\Delta_3}(U,V)\,=\, \frac{\pi^2}{2} \int \frac{ds}{4\pi i} \frac{dt}{4\pi i} &U^{\frac{s}{2}}
 V^{\frac{t}{2}-\frac{\Delta_1+\Delta_3}{2}} 
    \Gamma\left[\Delta_1-\frac{s}{2}\right] 
    \Gamma\left[\Delta_3-\frac{s}{2}\right] 
    \\
    &\times
    \left(\Gamma\left[\frac{\Delta_1+\Delta_3-t}{2}\right]
    \Gamma\left[\frac{\Delta_1+\Delta_3-u}{2}\right] 
    \right)^2\mathcal{M}_{\Delta_1,\:\!\Delta_3}(s,t)\,,
    \nonumber
\end{align}
where $u=2\Delta_1+2\Delta_3-s-t$.

In this section we focus on the all-light limit. As described in the Introduction and in Section~\ref{sec:probe-calc}, the order $\alpha^2$ gravity correlator that we have derived should encode information about the LLLL correlator of two chiral primary operators $\mathcal{O}_2$ with $\Delta_1=2$ and two descendants $\mathcal{D}_{k}$ with $\Delta_3=k+4$ in the supergravity regime,
\begin{align}
    \mathcal{C}^{(\mathrm{\sst{desc}})}_{2,k+4}(U,V) \ &{} \equiv \, \lim_{|\vec{x}_2| \to\infty} |\vec{x}_2|^4 \langle \mathcal{O}_2 (x_1) \bar{\mathcal{O}}_2 (x_2) \bar{\mathcal{D}}_{k} (x_3) \mathcal{D}_{k} (x_4) \rangle \Big|_{\mathrm{dyna}}     \nonumber\\
    &{} \,\equiv\,  \lim_{|\vec{x}_2| \to\infty} |\vec{x}_2|^4 \frac{1}{ (x_{12}^2)^2 (x_{34}^2)^{k+4}} \mathcal{G}^{(\mathrm{\sst{desc}})}_{2,k+4}(U,V) \,=\, \frac{1}{U^{k+4}} \mathcal{G}^{(\mathrm{\sst{desc}})}_{2,k+4}(U,V)  \,. 
\end{align}

We now take the Mellin transform of the four-point correlator derived from supergravity, expressed in terms of the functions $\hat{D}_{\Delta_1 \Delta_2 \Delta_3 \Delta_4}(U,V)$.
The Mellin transform of the functions $\hat{D}_{\Delta_1 \Delta_2 \Delta_3 \Delta_4}(U,V)$ is given by
\begin{align}
    \hat{D}_{\Delta_1 \Delta_2 \Delta_3 \Delta_4}(U,V) \;=\;\:
    &
    \frac{\pi^{2}}{2\prod\limits_{j=1}^4 
    \Gamma(\Delta_j)} \Gamma\bigg(\frac{\hat{\Delta}-4}{2}\bigg)
    \int \frac{ds}{4\pi i} \frac{dt}{4\pi i} U^{\frac{s}{2}} V^{\frac{t}{2}} \Gamma\left[-\frac{s}{2}\right] \Gamma\left[-\frac{t}{2}\right] 
    \cr
    &~~~
    \times
    \Gamma\left[\Delta_4+\frac{s+t}{2}\right] 
    \Gamma\left[ \frac{\Delta_1+\Delta_2 -\Delta_3 - \Delta_4 -s}{2} \right] 
    \\
    &~~~
    \times
    \Gamma\left[ \frac{\Delta_2+\Delta_3 -\Delta_1 - \Delta_4 -t}{2} \right] \Gamma\left[ \frac{\Delta_1+\Delta_3 +\Delta_4 - \Delta_2 + s +t }{2} \right]\,.~~~~~
    \nonumber
\end{align}

Then up to numerical factors that do not depend on $k$, we obtain the following result for the LLLL correlator of two CPOs and two descendants: 
\begin{align}
    \mathcal{M}_{2,k+4} \,=\; &{}\frac{1}{d_k}\left(\frac{\left(k^2+7 k+12\right) u^2-2 \left(k^3+10 k^2+41 k+60\right) u+k^4+13 k^3+78 k^2+240 k+304}{s-2}\right.~~ \nonumber\\
    \label{eq:sugra-mellin}
    &{}\qquad \left.+\frac{8 k (k+1)}{u-(k+4)}+ (k+3)\left( (k+4) u -k^2-2 k-16\right)\right) ,
\end{align}
where $d_k=\Gamma(k+4)$ and $u=2k+12-s-t$.

\subsection{Superconformal Ward identity}

The above set of all-light correlators of two CPOs and two descendants is related by a superconformal Ward identity to a set of correlators of four CPOs that were computed in tree-level supergravity for $k=0$ in~\cite{Arutyunov:2000py,Uruchurtu:2008kp}.
We now verify that this Ward identity is satisfied.

We continue to focus on the dynamical part of the correlator in the supergravity regime. We thus introduce the notation
\begin{align}
     {\mathcal{C}}^{(\mathrm{\sst{CPO}})}_{2,k+2}(U,V) &{} \:\equiv \, \lim_{|\vec{x}_2| \to\infty} |\vec{x}_2|^4 \langle \mathcal{O}_2 (x_1) \bar{\mathcal{O}}_2 (x_2) \bar{\mathcal{O}}_{k+2} (x_3) \mathcal{O}_{k+2} (x_4)  \rangle 
     \Big|_{\mathrm{dyna}}
     \nonumber\\
    &{} \,\equiv\, \lim_{|\vec{x}_2| \to\infty} |\vec{x}_2|^4 \frac{1}{ (x_{12}^2)^2 (x_{34}^2)^{k+2}} \,  {\mathcal{G}}^{(\mathrm{\sst{CPO}})}_{2,k+2}(U,V) \,=\, \frac{1}{U^{k+2}} {\mathcal{G}}^{(\mathrm{\sst{CPO}})}_{2,k+2}(U,V)  \,. 
\end{align} 
It is common to introduce an auxiliary complex null vector $\vec{y}$ to write the general single-trace operator in the form~\cite{Arutyunov:2002fh,Nirschl:2004pa,Dolan:2006ec}
\be
    \cO_p \,=\, y^{i_1} \ldots y^{i_p} \, \mathrm{Tr} \left( \phi_{i_1} \ldots \phi_{i_p} \right)\,, \qquad \vec{y}\cdot \vec{y} \,=\, 0\,.
\ee
Introducing the cross-ratio-like variables
\be
    \sigma \,=\, \frac{y_{13}^2 y_{24}^2}{y_{12}^2 y_{34}^2} \:, \qquad \tau \,=\, \frac{y_{14}^2 y_{23}^2}{y_{12}^2 y_{34}^2} \:, \qquad
    y_{ij}^2 \,\equiv\, \vec{y}_i \cdot \vec{y}_j \:,
\ee
the four-point correlators of interest take the form~\cite{Nirschl:2004pa}
\be 
    \langle \mathcal{O}_2 (x_1) \mathcal{O}_2 (x_2) \mathcal{O}_{p} (x_3) \mathcal{O}_{p} (x_4)  \rangle \,=\,
    \left(\frac{y_{12}^2}{x_{12}^2}\right)^2 \left(\frac{y_{34}^2}{x_{34}^2}\right)^p
    {\mathcal{G}}^{(\mathrm{\sst{CPO}})}_{2 2 p p}(U,V,\sigma,\tau)\,,
\ee
where 
\be
   {\mathcal{G}}^{(\mathrm{\sst{CPO}})}_{2 2 p p}(U,V,\sigma,\tau) \,=\, \mathcal{I}(U,V,\sigma,\tau) {\mathcal{H}}^{(\mathrm{\sst{CPO}})}_{2 2 p p}(U,V,\sigma,\tau)\,,
\ee
and
\be
   \mathcal{I}(U,V,\sigma,\tau) \,=\, V + \sigma ^2 U V + \tau ^2 U + \sigma  V (V-U-1) +\tau  (1-U-V)  + \sigma  \tau  U (1-U-V)\,. 
\ee

For our choice of operators ($\mathcal{O}_p = \mathrm{Tr} \, Z^p$, $\,p=k+2$, $\,Z = \phi_1+i\phi_2$) we have
\be
   \vec{y}_1 \,=\, \vec{y}_4 \,=\, \left(1, i,0,0,0,0\right)\,, \qquad \vec{y}_2 \,=\, \vec{y}_3 \,=\, \left(1,-i,0,0,0,0\right)\,,
   \quad~\Rightarrow\quad~ \sigma \,=\,1\,, \quad \tau \,=\, 0
   \,,
\ee
so $\mathcal{I}=V^2$. The dynamical part of the 22$pp$ correlator in the supergravity regime is given by (see e.g.~\cite{Aprile:2017xsp})
\be
    {\mathcal{H}}^{(\mathrm{\sst{CPO}})}_{2 2 p p} \:=\: U^p \, \bar{D}_{p,\, p+2,\, 2,\, 2} \;.  
\ee
The superconformal Ward identity involves the following differential operator \cite{Drummond:2006by,Goncalves:2014ffa}
\be
    \Delta^{(2)} \,=\, U\partial_U^2 + V\partial_V^2 + \left(U+V-1\right)\partial_U\partial_V + 2\left(\partial_U+\partial_V\right)\,,
\ee
and takes the form
\be
\label{eq:WI}
    \mathcal{C}^{(\mathrm{\sst{desc}})}_{2,k+4}(U,V) \,=\, \left(\Delta^{(2)}\right)^2 {\mathcal{C}}^{(\mathrm{\sst{CPO}})}_{2,k+2}(U,V) \:.
\ee
Using the following properties of the $\bar{D}$-functions~\cite{Gary:2009ae},
\begin{align}
\begin{aligned}
\partial_U \bar{D}_{\Delta_1,\Delta_2,\Delta_3,\Delta_4}(U,V) \;&=\;    
-\bar{D}_{\Delta_1+1,\Delta_2+1,\Delta_3,\Delta_4}(U,V) \,, \\
\partial_V \bar{D}_{\Delta_1,\Delta_2,\Delta_3,\Delta_4}(U,V) \;&=\;    
-\bar{D}_{\Delta_1,\Delta_2+1,\Delta_3+1,\Delta_4}(U,V) \,,
\end{aligned}
\end{align}
we take the Mellin transform of the RHS of the Ward identity~\eqref{eq:WI}. When comparing this to the result of our supergravity calculation, the coefficient of the pole at $s=2$ determines the overall normalization. We then find precise agreement with the supergravity result for $\mathcal{M}_{2,k+4}$ in Eq.~\eqref{eq:sugra-mellin}.

\section{Discussion}
\label{sec:disc}

In this paper we have reported a new computation of four-point correlation functions of two chiral primary operators and two descendants in the supergravity regime, by computing holographic two-point functions in non-trivial LLM supergravity solutions. The family of LLM solutions we studied are those specified by a profile with the shape of a ripple on a circle. The light probe in supergravity was taken to be an infinite sequence of Kaluza-Klein harmonics of the dilaton/axion.

This work represents a novel application of the prescription for computing heavy-heavy-light-light four-point correlators to AdS$_5$/CFT$_4$ holography. 
While the LLM solutions we have studied are generically heavy, and the method we use is a priori for HHLL correlators, we have worked perturbatively in $\alpha$ and our main focus has been the all-light limit of the supergravity calculation.
The advantage of using this method to compute LLLL correlators is that it avoids the use of Witten diagrams and their associated complications.

The all-light correlators we have computed are related by superconformal Ward identities to previously known four-point correlators of CPOs. We verified that the Ward identities are indeed respected, thus confirming the validity of the supergravity method.

In similar investigations of LLLL four-point correlators in AdS$_3$ by probing non-trivial smooth supergravity solutions~\cite{Giusto:2018ovt,Giusto:2019pxc,Giusto:2020neo}, oftentimes (but not always)
the supergravity calculation directly yielded only a subset of terms of the LLLL correlators (those corresponding to exchange of single-trace operators in the channel in which the two light operators become close). The remaining parts of the correlators were reconstructed by consistency considerations. We have seen that in the present work, the light limit of the supergravity calculation yields the full dynamical part of the LLLL correlators.

It is interesting to reflect that the computation of the correlators of four CPOs of the form $\langle \cO_2 \cO_2 \cO_p \cO_p \rangle$ for general $p$ in the supergravity regime via Witten diagram methods took several years of effort~\cite{Arutyunov:2000py,Arutyunov:2002fh,Berdichevsky:2007xd,Uruchurtu:2008kp}. With the benefit of hindsight, the calculation we have presented in this work could in principle have been used as a way of obtaining information about four-point correlators of CPOs, via the same Ward identities we have studied, at an earlier stage in this endeavour.

Our computation raises interesting questions for future work. It would be very interesting to understand some more details of the family of CFT heavy states dual to the LLM solutions we have studied. This can be done using precision holography, by extending and refining the analysis of~\cite{Skenderis:2006uy,Skenderis:2007yb}, taking into account the observations of~\cite{Giusto:2024trt}, analogously to the recent developments in precision holography in AdS$_3\times $S$^3$~\cite{Giusto:2015dfa,Giusto:2019qig,Rawash:2021pik}.
In addition, our methods can be generalized to compute more general families of correlators similar to those derived here. Work in both of these directions is in progress, and we intend to report further on these questions in due course.

Looking further to the future, it is well known that supersymmetric asymptotically AdS$_5$ black holes with finite horizons in supergravity are 1/16-BPS~\cite{Gutowski:2004yv}. A long-standing open problem is to ask whether one can understand (some or all of) the microstates of this black hole in the bulk description. Interpolating between the 1/2-BPS and 1/16-BPS sectors, there are 1/4 and 1/8-BPS generalizations of the LLM solutions, see~\cite{Chen:2007du,Lunin:2008tf,Jia:2023iaa} and refs within. 
There are many interesting open questions related to supersymmetric black holes and their microstates,
and there is much scope for further work on the description of heavy pure states in both bulk and boundary theories, and on correlators involving probes of such pure states.

\vspace{3mm}

\section*{Acknowledgements}

For valuable discussions, we thank Stefano Giusto, Hynek Paul, Rodolfo Russo, Kostas Skenderis, Mritunjay Verma, and Congkao Wen. The work of DT was supported by a Royal Society Tata University Research Fellowship. The work of A.T.~was supported by a Royal Society Research Fellows Enhanced Research Expenses grant.

\vspace{4mm}

\begin{appendix}

\section{Metric at second order}
\label{app:metric2}

After the second-order gauge transformation generated by $\xi^{(2)}$, given in Eq.~\eqref{eq:gauge-2}, the second-order metric $g^{(2)}$ is given by the following main components, with the remaining non-zero components following from $SO(4)\times SO(4)$ symmetry. The order $\alpha^0$ line element on the three-sphere in S$^5$ (c.f.~Eq.~\eqref{eq:metric0}) is taken to be $\,d{\Omega}_3^2\,=\;\!
d\eta^2+\sin^2\eta(d\chi^2+\sin^2\chi d\psi^2)\;\!$.
%
\begin{align}
    g^{(2)}_{tt} &\,=\, \frac{1}{128 \left(R^2+1\right)^3}\Bigg(
    96 R^8+304 R^6+331 R^4+238 R^2+43  
    \cr
    &{}\qquad\qquad\qquad\qquad~~~
    -4 \left(12 R^6+29 R^4+58 R^2+1\right) \cos 2 \theta \\[3mm]
    &{}\qquad\qquad\qquad\qquad~~~
    +\left(13 R^4+38 R^2-7\right) \cos 4 \theta \cr
    &{}\qquad\qquad\qquad\qquad~~~
    -8 \left(R^2+1\right) \left(7 R^2+11\right) \sin^4\theta \cos \!\!\:\big(4 (t-\phi)\big)\Bigg)
    \,, \qquad\quad~~
    \nonumber
    %
    %
    \end{align}
    \begin{align}
    g^{(2)}_{RR} 
    &\,=\, -\frac{1}{{128 \left(R^2+1\right)^5}}\Bigg(
    80 R^6+179 R^4+486 R^2+27 \cr
    &{}\qquad\qquad\qquad\qquad~~~~
    + 4 \left(12 R^6+71 R^4+38 R^2-13\right) \cos 2 \theta \\[3mm]
    &{}\qquad\qquad\qquad\qquad~~~~
    +\left(-31 R^4+38 R^2+5\right) \cos 4 \theta
    \cr
    &{}\qquad\qquad\qquad\qquad~~~~
    -8 \left(R^2+1\right) \left(11 R^2+7\right) \sin ^4\theta \cos \!\!\:\big(4 (t-\phi)\big)\Bigg) \,,
    \qquad\quad~~~
    \nonumber
    %
    %
    \end{align}
    \begin{align}
    g^{(2)}_{\beta\beta} &\,=\, \frac{R^2}{128 \left(R^2+1\right)^3} \Bigg(-96 R^6-304 R^4-283 R^2-27 
        \cr
    &{}\qquad\qquad\qquad\qquad~~
    + 4 \left(R^2+1\right) \left(12 R^2+13\right) \cos 2 \theta -\left(13 R^2+5\right) \cos 4 \theta
        \\
    &{}\qquad\qquad\qquad\qquad~~
    +56 \left(R^2+1\right) \sin ^4\theta \cos \!\!\:\big(4 (t-\phi)\big)\Bigg)\,,
    \cr
    g^{(2)}_{t R}
    &\,=\,\frac{R \sin ^4\theta  \sin \!\!\:\big(4(t-\phi )\big)}{4 \left(R^2+1\right)^3}\,,\\
    g^{(2)}_{\theta \theta}
    &\,=\,\frac{1}{384 \left(R^2+1\right)^3}
    \Bigg( 
    48 R^4+135 R^2-81 -4 \left(36 R^4+105 R^2+29\right) \cos 2 \theta 
      \\[-2mm]
    &{}\qquad\qquad\qquad\qquad~~
    +\left(93 R^2+37\right) \cos 4 \theta +264 \left(R^2+1\right) \sin ^4\theta  \cos \!\!\:\big(4 (t-\phi)\big)\Bigg)\,, \hspace{-3mm}
    \nonumber
    %
    \end{align}
    \begin{align}
     g^{(2)}_{\phi \phi}
     &\,=\,\frac{\sin ^2\theta }{384 \left(R^2+1\right)^3}
     \Bigg(
      48 R^4+135 R^2+15 -4 \left(36 R^4+105 R^2+53\right) \cos 2 \theta 
       \\[-2mm]
    &{}\qquad\qquad\qquad\qquad~~
    +\left(93 R^2+37\right) \cos 4 \theta +264 \left(R^2+1\right) \sin ^4\theta \cos \!\!\:\big(4 (t-\phi)\big)\Bigg)\,,
    \nonumber
    %
    \end{align}
    \begin{align}
    ~~~~
    g^{(2)}_{\eta \eta}&\,=\,\frac{\cos ^2\theta}{384 \left(R^2+1\right)^3} \Bigg(48 R^4+135 R^2-49 -12 \left(12 R^4+35 R^2+7\right) \cos 2 \theta 
      \\[-2mm]
    &{}\qquad\qquad\qquad\qquad~~
    +\left(93 R^2+37\right) \cos 4 \theta +264 \left(R^2+1\right) \sin ^4\theta  \cos \!\!\:\big(4 (t-\phi)\big)\Bigg)\,,~~~~
    \nonumber\\
      g^{(2)}_{t \phi}
      &\,=\,
      \frac{\sin ^2\theta  }{4 \left(R^2+1\right)^3}
      \bigg(\!-2 R^4-6 R^2-1 +\left(4 R^2+1\right) \cos 2 \theta \bigg)\,.
\end{align}

\section{Source terms of the second-order equation}
\label{app:sources}

\subsection{Projected sources in hyperspherical coordinates}

In this appendix we record the results of the projection of the sources~\eqref{eq:sources-defn} onto the appropriate spherical harmonic, as described around Eq.~\eqref{eq:proj-sources}. For $\big\langle\mathcal{J}_1\big\rangle$, we obtain
\begin{align}
     \big\langle\mathcal{J}_1\big\rangle \;\!=~& \frac{1}{d^{(1)}_k(R)} \Bigg\{ k \left(R^2+1\right) \biggl[ 
     24 (k+3) (k+4)^2 R^8 
     +32 (k+4)^2 (4 k+9) R^6
     \nonumber\\   &{} 
     \qquad\qquad\qquad\qquad\qquad
     +(k+4) \Big(257 k^2+1287 k+1564\Big) R^4
     \nonumber\\   &{} 
     \qquad\qquad\qquad\qquad\qquad~~
     +4 (k+4) 
     \Big( 49k^2 +183k+148 \Big) R^2
    \nonumber\\  &{}
    \qquad\qquad\qquad\qquad\qquad~~~~~
    + 43 k^3+249 k^2 -376 k-1584 \biggr] \hat{B}_{k+4} 
    \nonumber\\
    &{}
      \qquad\quad~
      - R \left(R^2+1\right) 
      \biggl[
      120 (k+3) (k+4) R^8
      +160 (k+4) (4 k+9) R^6
      \nonumber\\
      &{}
      \qquad\qquad\qquad\qquad\qquad
      +\Big(1417 k^2+6975k+7868\Big) R^4
      \nonumber\\
      &{}
      \qquad\qquad\qquad\qquad\qquad~~
      +4 \Big(393 k^2+1391k+836\Big) R^2
      \\
      &{}      \qquad\qquad\qquad\qquad\qquad~~~~~
      + 1287 k^2+3153 k-924
      \biggr] 
      \partial_R\hat{B}_{k+4}  
      \nonumber\\
      &{}
      \qquad\quad
      - R^2 \left(R^2+1\right)^2  \biggl[
      24 (k+3) (k+4) R^6
      +8 (k+4) (13 k+27) R^4
      \nonumber\\
      &{}
      \qquad\qquad\qquad\qquad\qquad
       +\Big(221 k^2+939k+748\Big) R^2
       +315 k^2+829 k-236\biggr] \partial^2_R\hat{B}_{k+4}
      \nonumber\\
      &{} 
      \qquad\quad
      + 32 i k(k+4) \left(R^2+1\right)  \biggl[
      (k+3) R^4+(5 k+7) R^2+k+1
      \biggr] \partial_t\hat{B}_{k+4}
      \nonumber\\
      &{} 
      \qquad\quad
      + \biggl[24 (k+3) (k+4) R^6+8 (k+4) (9 k+23) R^4
      \nonumber\\
      &{}
      \qquad\qquad\qquad
      +\Big(281 k^2+975k+892\Big) R^2
      +11 k^2-19 k-108\biggr] \partial^2_t\hat{B}_{k+4}  \Bigg\}
     \,,
     \nonumber
\end{align}
where the denominator $d^{(1)}_k(R)$ is given by
\be
d^{(1)}_k(R) \,=\, 32 (k + 3) (k + 4) (R^2 + 1)^5 \,.
\ee
For $\big\langle\mathcal{J}_2\big\rangle$, we obtain
\begin{align}
     \big\langle\mathcal{J}_2 \big\rangle \;=\;& \;
     \frac{1}{d^{(2)}_k(R)} \Bigg\{ 
      - 3k \left(R^2+1\right) 
     \biggl[ 
     (k+4)\Big(k^2-k+6\Big)R^4
     +2\Big(k^3 + 3  k^2 + 10 k + 32\Big) R^2
     \nonumber\\   &{} 
     \qquad\qquad\qquad\qquad\qquad~~~
     + k^3 + 3  k^2 - 30 k - 8
     \biggr] 
     \hat{B}_{k+4}
     \nonumber\\   &{} 
     \qquad\qquad~~~
     - 3(k+1) R (R^2 + 1) 
     \biggl[
     k  \big(k^2 + 6k - 2\big) R^4 
     + 2k \big(k^2 + 6  k - 10\big) R^2 
     \nonumber\\   &{} 
     \qquad\qquad\qquad\qquad
     \qquad\qquad\qquad\quad~~
     + k^3 + 6k^2 - 42  k - 48  
     \biggr]
     \partial_R \hat{B}_{k+4}
     \nonumber\\   &{} 
     \qquad\qquad\quad
     +12(k + 1) R^2 (R^2 + 1)^2 
     \Big[
     (2 k + 3) R^2 + 4 k + 7  
     \Big]
     \partial_R^2 \hat{B}_{k+4}
     \nonumber\\   &{} 
     \qquad\qquad\quad
     +3 (k + 1) (k + 2) R^3 (R^2 + 1)^3
     \partial_R^3 \hat{B}_{k+4}
     \nonumber\\   &{} 
     \qquad\qquad\quad
     - 6i k (k + 1) (k + 4) (R^2 + 1) 
     (3 R^2 - 2)\partial_t \hat{B}_{k+4}
     \nonumber\\   &{} 
     \qquad\qquad\quad
     + 6 i k  (k + 1)  (k + 4)  R (R^2 + 1)^2 \partial_t\partial_R \hat{B}_{k+4}
     \\   &{} 
     \qquad\qquad\quad
     -3 (k + 1) \Big[  (11  k + 18)  R^2 - 3  k - 10 \Big]
     \partial_t^2 \hat{B}_{k+4}
     \nonumber
     \\   &{} 
     \qquad\qquad\quad
     +3 (k + 1) (k + 2) R (R^2 + 1)
     \partial_t^2 \partial_R \hat{B}_{k+4}
     \Bigg\}
     \nonumber\\   &{}
     +\frac{3 k \;\! e^{i t}}{d^{(3)}_k(R)}
     \Bigg\{ 
     \Big[ k (k + 3)(R^2 + 1)^2 - 8 
     \Big]\hat{B}_{k+3}
     -4R (R^2 + 1)(R^2 + 2)
     \partial_R \hat{B}_{k+3}
     \nonumber\\   &{} 
     \qquad\qquad\quad~~
     -R^2 (1 + R^2)^2  \partial_R^2 \hat{B}_{k+3} + i(R^2-5)\partial_t \hat{B}_{k+3}
     \nonumber\\   &{} 
     \qquad\qquad\quad~~
    - 2i R (1 + R^2)\partial_t \partial_R \hat{B}_{k+3} +\partial_t^2 \hat{B}_{k+3} 
     \Bigg\} 
     \;,
     \nonumber
\end{align}
where
\begin{align}
d^{(2)}_k(R) \,&=\, 4 (k + 3) (k + 4) (R^2 + 1)^5 \,, \\
d^{(3)}_k(R) \,&=\, (k + 3)^2 (R^2 + 1)^{7/2} \,.
\end{align}


\newpage 

\subsection{Dynamical part of the source of the second-order equation}
\label{app:source-B}

The dynamical part of the source of the order $\alpha^2$ equation, defined around Eq.\;\eqref{eq:j-dyna}, written in terms of the functions $B^{\pm}_\Delta$, $\hat{B}_\Delta$ defined in Eq.~\eqref{eq:b-to-bdy-hy}, takes the following form. The spacetime dependence is suppressed throughout.
\begin{align}
   \!\!\!
   \mathcal{J}
   \;=&\; \frac{12  k}{k + 3} 
   \left(
   B^{-}_2 B^{+}_3 
   - \frac{2}{(k + 3)}
   B^{-}_3 B^{+}_4
   \right)\hat{B}_{k+3}
   \nonumber\\ &
   + 12  k B^{-}_1 B^{+}_3 \hat{B}_{k+4} 
   - \frac{17 k^3 + 123 k^2 + 328 k + 48}{8 (k + 3)}
   B^{-}_2 B^{+}_2 \hat{B}_{k+4}
   \nonumber\\ &
   +\frac{\left(-3 k^4-14 k^3+281 k^2+2216 k+3088\right)}{8 (k+3) (k+4)} B^{-}_3 B^{+}_3 \hat{B}_{k+4}
   \nonumber\\ &
   -\frac{24 k}{k+3} B^{-}_2 B^{+}_4 \hat{B}_{k+4}
   +\frac{9 \left(k^3+11 k^2-88 k-208\right)}{8 (k+3) (k+4)} B^{-}_4 B^{+}_4 \hat{B}_{k+4}
   \nonumber\\[2mm] &
   +k(k+4) \Big( 
   B^{-}_2 B^{+}_1 +
   3 B^{-}_1 B^{+}_2
   \Big)
   \hat{B}_{k+5}
   \nonumber\\ &
   -\frac{12 k (k+4) }{k+3}
   B^{-}_1 B^{+}_4 \hat{B}_{k+5}
   \nonumber\\ &
   +\frac{1}{16 (k+3)}
   \Big(109 k^3+874 k^2+2021 k+692\Big)
    B^{-}_2 B^{+}_3 \hat{B}_{k+5}
   \nonumber\\ &
   +\frac{1}{16 (k+3)}
   \Big(48 k^4+589 k^3+2330 k^2+3237 k+692\Big)
    B^{-}_3 B^{+}_2 \hat{B}_{k+5}
   \nonumber\\ &
   -\frac{9}{8 (k+3)}
   \Big(k^3+22 k^2+149 k+244\Big)
   B^{-}_3 B^{+}_4 \hat{B}_{k+5}
   \nonumber\\ &
   -\frac{3}{8} \Big(19 k^2+89 k+244\Big) B^{-}_4 B^{+}_3 \hat{B}_{k+5}
   %
   \label{eq:source-prop}
   \\ &
   +\frac{9 \left(k^2+7 k+20\right) }{2 (k+3)}
   \Big(
   B^{-}_4 B^{+}_5
   +B^{-}_5 B^{+}_4  
   \Big) 
   \hat{B}_{k+5}
   \nonumber\\ &
   -\frac{2}{k+3}
   \Big(k^3+10 k^2+29 k+20\Big)
   B^{-}_2 B^{+}_2 \hat{B}_{k+6}
   \nonumber\\ &
   -\frac{k^3+8 k^2+11 k-20}{k+3}
   \Big(
    B^{-}_1 B^{+}_3
    +
    B^{-}_3 B^{+}_1
    \Big)
    \hat{B}_{k+6}
   \nonumber\\ &
   -\frac{3}{16 (k+3)}
   \Big(25 k^3+236 k^2+599 k+220\Big)
   B^{-}_2 B^{+}_4 \hat{B}_{k+6}
   \nonumber\\ &
   -\frac{1}{4 (k+3)}
   \Big(24 k^4+329 k^3+1508 k^2+2551 k+1180\Big)
   B^{-}_3 B^{+}_3 \hat{B}_{k+6}
   \nonumber\\ &
   -\frac{3}{16 (k+3)}
   \Big(16 k^4+185 k^3+700 k^2+919 k+220\Big)
   B^{-}_4 B^{+}_2 \hat{B}_{k+6}
   \nonumber\\ &
   +\frac{9 }{4 (k+3)}
   \Big(5 k^3+44 k^2+123 k+140\Big)
   B^{-}_4 B^{+}_4 \hat{B}_{k+6}
   \nonumber\\ &
   +\frac{3 }{2 (k+3)}
   \left(
   k^3+12 k^2+55 k+100
   \right)
   \Big(
    B^{-}_3 B^{+}_5 
    +
    B^{-}_5 B^{+}_3
    \Big) 
    \hat{B}_{k+6}
   \nonumber\\ &
   +\frac{3}{k+3}
   \left(
   k^4+14 k^3+65 k^2+112 k+60
   \right)
   \Big(
   B^{-}_3 B^{+}_4
   +
   B^{-}_4 B^{+}_3 
   \Big) 
   \hat{B}_{k+7}~.
   \nonumber
\end{align}

\end{appendix}

\newpage

\begin{adjustwidth}{-3mm}{-3mm} 
\bibliographystyle{utphys}      
\bibliography{microstates}       

\end{adjustwidth}


\end{document}